\documentclass[prl,twocolumn,showpacs,preprintnumbers,amsmath,amssymb,superscriptaddress]{revtex4-1} 
\usepackage{graphicx}
\usepackage{bm}
\usepackage{dsfont} 
\usepackage{color} 
\usepackage{multirow}
\usepackage{dcolumn}
\usepackage{enumitem}

\usepackage{pstricks}% defines \newrgbcolor

\newrgbcolor{mygr}{.5 .5 .5}

\newrgbcolor{mydb}{0 0 .2}
\newrgbcolor{mydp}{.5 .5 .5}
\newrgbcolor{mydor}{.7 .4 0}

\newrgbcolor{colLM}{.05 .7 .4}
\newrgbcolor{colCM}{.05 .5 .5}
\newrgbcolor{colRM}{.2 .05 .7}

\newcommand{\bhvr}{behavior} 			% American English 

\newcommand{\Ekin}{E_\text{kin}}

\newcommand{\Happ}{H_\text{app}}

\newcommand{\Japp}{J_\text{app}}
 
\newcommand{\Keff}{K_\text{eff}}

\newcommand{\mgn}{magnetization} 
\newcommand{\Mgn}{Magnetization} 
\newcommand{\MS}{M_\text{S}} 
 
\newcommand{\ud}{\mathrm{d}}
\newcommand{\udt}{\frac{\mathrm{d}}{\mathrm{d} t }}

\newcounter{myCounter}
\begin{document} 

\topmargin 0.0cm  

\preprint{APS/123-QED} 

\title{Spin-orbit torque driven chiral magnetization reversal in ultrathin nanostructures}  

\author{N.~Mikuszeit}
\email{Nikolai.Mikuszeit@gmail.com} 
\author{O.~Boulle}
\author{I.~M.~Miron}
\affiliation{Univ.\ Grenoble Alpes, INAC-SPINTEC, F-38000 Grenoble, France} 
\affiliation{CNRS, INAC-SPINTEC, F-38000 Grenoble, France} 
\affiliation{CEA, INAC-SPINTEC, F-38000 Grenoble, France} 
\author{K.~Garello}
\author{P.~Gambardella}
\affiliation{Department of Materials, ETH Z{\"u}rich, H{\"o}nggerbergring 64, Z{\"u}rich CH-8093, Switzerland}
\author{G.~Gaudin}
\author{L.~D.~Buda-Prejbeanu}
\affiliation{Univ.\ Grenoble Alpes, INAC-SPINTEC, F-38000 Grenoble, France} 
\affiliation{CNRS, INAC-SPINTEC, F-38000 Grenoble, France} 
\affiliation{CEA, INAC-SPINTEC, F-38000 Grenoble, France}

\date{\today}			% It is always \today, today, 
             			%  but any date may be explicitly specified 
 
\begin{abstract} 
We show that the spin-orbit torque induced magnetization switching in nanomagnets presenting  
Dzyaloshinskii-Moriya (DMI) interaction is governed by a chiral 
domain nucleation at the edges. The nucleation is induced by the DMI and 
the applied in-plane magnetic 
field followed by domain wall propagation.  Our micromagnetic simulations show that the DC 
switching current can be defined as the edge nucleation current, which decreases strongly 
with increasing amplitude of the DMI. This description allows us to build a simple 
analytical model to 
quantitatively predict the switching current. 
We find that domain nucleation occurs down to a lateral size of $15\;$nm, 
defined by the length-scale of the DMI, beyond which the reversal 
mechanism approaches a macrospin behavior. The switching is 
deterministic and bipolar 
\end{abstract} 

%75.60.Jk 	Magnetization reversal mechanisms
%85.70.Ay 	Magnetic device characterization, design, and modeling
%85.75.Dd 	Magnetic memory using magnetic tunnel junctions
                          
% Use showkeys class option if 
% keyword display desired 
\maketitle

The recent discovery that a current can switch the magnetization of a nanomagnet 
in ultrathin heavy metal (HM)/Ferromagnetic (FM) multilayers has opened a new path to 
manipulate \mgn\ at the nanoscale~\cite{Miron_Nat2011}. 
The switching arises from structural inversion asymmetry and high spin-coupling, 
resulting in a %charge current induced
spin current from the HM into the FM. 
This novel switching mechanism has led to innovative magnetic memory concept, namely the 
spin-orbit torque MRAM~\cite{Miron_Nat2011,Liu_Science2012,Cubukcu_APL2014}, 
which combines large endurance, low power, and fast switching and thus 
appears as a possible non-volatile alternative 
for cache memory applications. %~\cite{Kim_12NM,Fan_13NC,Garello_13NN,Qiu_15NN}
Recently, Garello \emph{et al.}~\cite{Garello_APL2014} demonstrated deterministic 
magnetization switching by spin-orbit torque (SOT)
in ultrathin Pt/Co/AlO$_x$, as fast as $180\;$ps. These observations could not be explained within 
a simple macrospin approach, suggesting a magnetization reversal mechanism by domain nucleation and domain wall (DW) propagation. 
The failure of the macrospin approach for quantitative description is also underlined by the predicted switching current 
density, which is nearly one order of magnitude larger than experimental ones~\cite{Liu_PRL2012,Lee_APL2013,Lee_APL2014}. 
Besides its fundamental importance, this lack of a proper quantitative modeling  is an important issue for 
the design of logic and memory devices based on SOT switching, which have so far considered a 
macrospin description~\cite{Kim_IEEE-ED2015,Jabeur_EL2014,Jabeur_IEEE-TransMag2014,Wang_JPD2015}.
The final ingredient is the presence of antisymmetric exchange interaction, 
i.e.\ Dzyaloshinskii-Moriya interaction (DMI). This exchange tends to form states of non-collinear magnetization,
promoting homochiral N{\'e}el DW ~\cite{Chen_PRL2013,Thiaville_EPL2012,Tetienne_NatComm2015}.   
In the N{\'e}el configuration, a maximal SOT  is applied on the 
DW ~\cite{Thiaville_EPL2012,Emori_NatMater2013,Boulle_PRL2013,Ryu_NatNano2013}. 
This explains the large current induced DW velocity observed 
experimentally~\cite{Miron_Nat2011,Ryu_NatNano2013}.
Moreover, the DMI can result in significant magnetization tilting at the edges of 
magnetic structures, resulting, e.g., in asymmetric field induced domain nucleation~\cite{Rohart_PRB2013, Pizzini_PRL2014}.
The influence of the DMI on the  magnetization pattern during SOT switching  was recently pointed out 
in micromagnetic simulation studies~\cite{Perez_APL2014a,Finocchio_APL2013,Martinez_SR2015}, 
whereas recent experimental work~\cite{Lee_PRB2014} explained the SOT switching mechanism 
by the expansion of a magnetic bubble.
%   where an in-plane external magnetic field is required  to overcome the DW chirality imposed by the DMI. 

Here, using micromagnetic simulations and analytic modeling, 
we show that the SOT-induced magnetization switching in the presence of DMI is 
governed by domain nucleation on one edge followed by propagation to the opposite edge. 
This reversal process allows to explain the ultra-fast deterministic switching observed experimentally.  
We systematically demonstrate that DMI leads to a large decrease of
the switching current and of the switching time and thus strongly affects the reversal energy. 
On the basis of our micromagnetic simulations, we provide a simple analytical model which allows 
to quantitatively predict the SOT switching current in the presence of DMI. Finally, we address 
the evolution of the switching mechanism   as the lateral dimension decreases, which is a  key 
feature for the device scalability.

 	\begin{figure*}
		\includegraphics[width=0.85\linewidth]{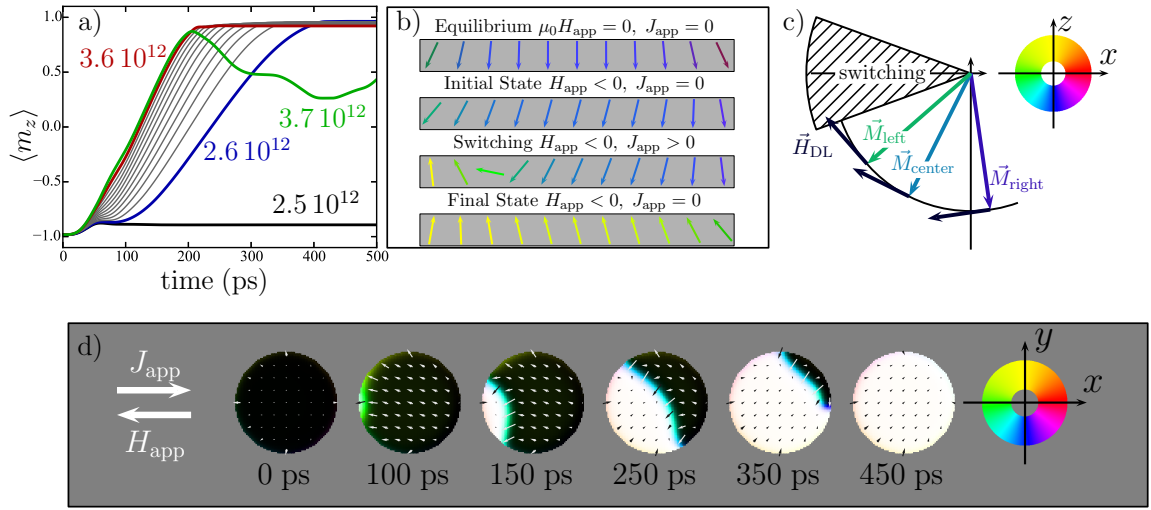}
		\caption{\label{fig:1}
			(color online)  a) Time evolution of the average out-of-plane magnetization 
			for different applied current densities %. % and $D = -2\;\mathrm{mJ/m}^2$. 
			%The current varies 
			(variations in steps of $10^{11}\;\mathrm{A/m}^2$). 
			The minimum current to  trigger switching, i.e.\  the critical current, 
			is highlighted in blue. The green curve indicates the threshold of 
			turbulent behavior (see text).
			b) Sketch of the magnetization configuration at different stages of the 
			switching process.
			c) \Mgn\ orientation in the center and at the left and right edges.
			The current induced damping-like torque (represented
			as effective field \textcolor{mydb}{$\vec H_\text{DL}$}) can only drive the left edge \mgn\ into instability, 
			resulting in a nucleation at the left edge.
			d) Snap-shots of the magnetization configuration showing the reversal from down (black) to
			up (white) via domain wall nucleation and propagation under an externally applied field of $\mu_0 H=0.1\;$T 
			and a current density of $2.6\;10^{12}\;\mathrm{A/m}^2$.
		}
	\end{figure*} 
The structures considered in this study are similar to the one used 
by Garello \emph{et al.}~\cite{Garello_APL2014}: a perpendicularly magnetized Co circular nanodot 
on top of a Pt stripe and capped with alumina. 
The DMI is  included into the simulation using the expression of Ref.~\cite{Thiaville_EPL2012}.
In addition to the standard micromagnetic energy density (which includes the exchange, the 
magnetocrystalline anisotropy, the Zeeman and the demagnetizing energy), 
the current injected in the Pt layer leads to two SOT terms in the
the Landau-Lifshitz-Gilbert equation: the  field-like 
 %$T_\text{FL}=\gamma_0 T_{FL} J \vec m \times \vec{e}_y$ 
$T_\text{FL}\propto \vec m \times \vec{e}_y$ 
and the  damping-like 
 %$T_\text{DL}=\gamma_0 T_{DL} \vec m \times (\vec m \times \vec{e}_y) $, 
$T_\text{DL} \propto \vec m \times (\vec m \times \vec{e}_y) $, 
where $\vec{e}_y$ is the unit vector in $y$-direction 
(see~\cite{spotSuppMat} for additional details). %,  
%$J$ is the current density, $ T_{FL}$ and  $T_{DL}$ describe the amplitude of the 
%field-like and damping like torque respectively, and $\gamma_0=\gamma/mu_0$ where 
%$\gamma$ is the gyromagnetic ratio. 
If not mentioned otherwise the external applied field is $\mu_0 \Happ=-0.1\;$T, 
and the material parameters 
are~\cite{Garello_Nat2013}: the saturation magnetization $M_\mathrm{S}=1090\;$kA/m, 
the exchange constant	$A_\text{ex}=1.0\times 10^{-11}\;$A/m, 
the perpendicular magnetic anisotropy constant $K_\text{u}=1248\;\mathrm{kJ}/\mathrm{m}^3$, 
the DMI amplitude $D=2\;\mathrm{mJ}/\mathrm{m}^2$, the Gilbert damping parameter 
$\alpha=0.5$, the torques $T^0_\text{FL}=-0.05\;\text{pT}\text{m}^2/\text{A}$ and 
$T^0_\text{DL}=+0.1\;\text{pT}\text{m}^2/\text{A}$.

The 3D micromagnetic simulations are performed using the solver Micro3D~\cite{Buda_CMS2002} with a 
mesh size smaller than $1.5\;$nm. The initial magnetization state of the dot is 
the remanent state after saturation by a negative magnetic field ($-O_z$) as shown for $0\;$ps in Fig.~\ref{fig:1} d). 
In the presence of an applied magnetic field $\vec \Happ$ in the $x$-direction,
magnetization dynamics is induced by a current pulse with a rise (and fall) time of $50\;$ps and variable width and amplitude.
Typical simulation results of a $100\;$nm dot are presented in Fig.~\ref{fig:1} a).
Depending on the current amplitude, three regimes are identified:

\noindent
\stepcounter{myCounter}
\themyCounter) For $ J_\text{app}  \leq 2.50\;10^{12}\;\mathrm{A/m}^2$
			no magnetization switching is observed. 
			The SOT leads to a slight tilting of the magnetization toward the plane of the dot, 
			but the magnetization relaxes toward its initial equilibrium state after the pulse.
			
\noindent
\stepcounter{myCounter}
\themyCounter) At intermediate current values 
			($2.60\;10^{12}\;\mathrm{A/m}^2 \leq  J_\text{app}  \leq 3.70\;10^{12}\;\mathrm{A/m}^2$)
			magnetization reversal occurs. The time evolution of the magnetization pattern in the dot, 
			see Fig.~\ref{fig:1} d) % for $ J_\text{app}   = 2.60\;10^{12}\;\mathrm{A/m}^2$) 
			reveals that, in contrast to recent interpretations~\cite{Lee_PRB2014}, the magnetization 
			reversal occurs by domain nucleation shortly after the pulse injection ($100\;$ps), 
			followed by fast DW propagation. 
			The nucleation always occurs on the left edge of the dot. 
			Once nucleated, the DW  propagates fast through the dot 
			and  is expelled on the opposed edge.   
			The switching  time $t_0$, defined by $\langle m_z\rangle(t=t_0)=0$, 
			decreases as $\Japp$ increases; the increase of the slope of $\langle m_z\rangle(t)$ 
			indicates that this is related to a faster DW propagation.
			As expected the DW has a N{\'e}el configuration due to the large DMI. 
			The simulation highlights that the DW nucleation occurs for all current values on the 
			same edge in a deterministic way. Symmetrically, when reversing the sign of the current, 
			the reversal from the up to the down state occurs on the opposite edge, 
			i.e.\ the behavior is bipolar.

\noindent
\stepcounter{myCounter}
\themyCounter) For higher currents ($J_\text{app} \geq 3.70\;10^{12}\;\mathrm{A/m}^2$) 
			the motion of the DW becomes turbulent (oscillatory) and the coherence of 
			the switching is destroyed.

%%%%%%%%%%% Simple Explanation of the physics   %%%%%%%%%%%%%%%%%%%%%%%%%
%
	\begin{figure*}
		\includegraphics[width=0.98\linewidth]{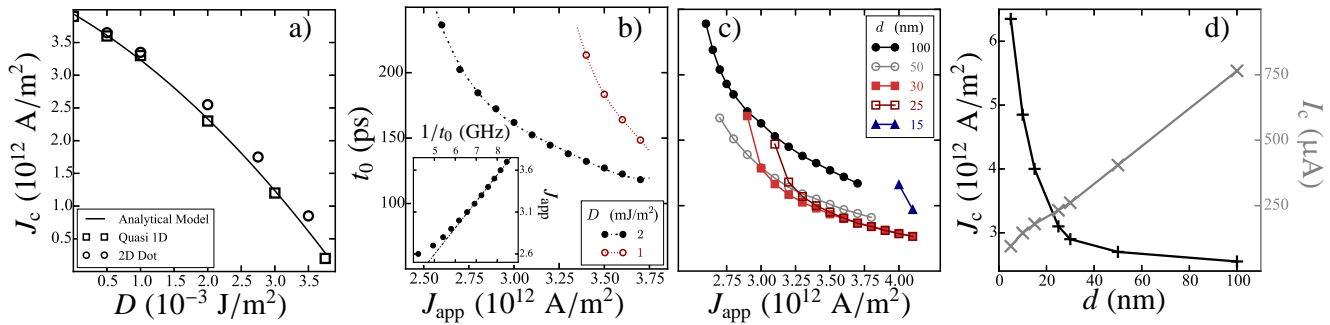}
		\caption{\label{fig:2} (color online)
		a) Critical current for destabilizing the system as a function of DMI strength. 
		b) The relation between the critical
		current and the switching time $t_0$ for two different values of DMI.The inset show the data
		for $D=2\;\text{mJ/m}^2$, but in a transformed coordinate system $\Japp$ verus $t_0^{-1}$.
		c) The switching time versus current for different dot
		diameters. 
		d) Critical current and current density for different dot sizes. 
		The calculation of the current assumes a 3 nm thick Pt line.
%		a) An increase of DMI strongly decreases the critical current $J_\mathrm{c}$ for switching and
%		b)the switching time $t_0$.
%		c) The switching time also decreases with decreasing dot diameter. Despite the switching mechanism changing near
%		the exchange length, 
%		d) the current show a linear and, therefore, scalable \bhvr.	
		}
	\end{figure*} 
The magnetization reversal scheme can be explained in a simplified manner by considering 
the combined effect of DMI, external magnetic field, and SOT, but neglecting small variations of the
demagnetizing field~\cite{Meckler_PRB2012}. % on the edge magnetization. %%impression that SOT only acts on edge
The DMI is too small to introduce a spin spiral but results in a magnetization canting at the dot 
edges~\cite{Rohart_PRB2013, Pizzini_PRL2014,Martinez_SR2015}. 
The edge canting can be considered as an additional effective field 
with spacial variation: on one side this field adds to the in-plane applied field, while it counteracts on the other 
(see Fig.~\ref{fig:1} b). This leads to an asymmetric tilting of the magnetization  on both edges. 

Upon current injection the damping-like torque emerges. Its effect can be interpreted as a %perpendicular 
rotating magnetic field of the form $\vec H_\text{DL}\propto J_\text{app}  \vec m \times\vec e_\mathrm{y}$ 
(see Fig.~\ref{fig:1} c)). 
This leads to a rotation of the magnetization towards the film plane 
on one side and away from the film plane on the other. Naturally, the current polarity is chosen such that the 
stronger tilted edge \mgn\ turns towards the film plane. 
Above a critical current an instability occurs, leading to domain nucleation and consecutive DW propagation.
It is clear that the current $J_\mathrm{c}$, required to introduce the instability, reduces with increasing DMI. 
This \bhvr\ is seen in Fig.\ref{fig:2} a), where $J_\mathrm{c}$ tends to zero when $D \approx 3.8\;\mathrm{mJ/m}^2$.
Moreover, for $\Japp>J_\mathrm{c}$ an increase of DMI decreases the switching time, 
as can be seen from Fig.~\ref{fig:2} b). 
After expelling the DW on the opposite side, switching has occurred and the more tilted edge appears on the opposite side.
As the SOT rotates this side away from the film plane and is not sufficient to rotate the less tilted side into instability,
the state is, hence, stable. 
It can be easily checked that this reversal scheme is in agreement with the hysteretic bipolar switching 
observed experimentally when sweeping $J_\text{app}$ and $H_\text{app}$~\cite{Miron_Nat2011}.

%%%%%%%%%%%%%%%%%    Model description %%%%%%%%%%%%%%
To understand these results better, we consider a simple analytical model, 
which describes the reversal process in the presence of both DMI and SOT. 
Using a Lagrangian approach and following Pizzini \emph{et al.}~\cite{Pizzini_PRL2014}, 
the strategy is, eventually, similar to the Stoner-Wohlfarth approach in a single domain particle but 
using the energy functional
	\begin{equation}
		\label{eq:et}
		\frac{E(\theta)}{V}=-\Keff\cos^2\theta-  \MS \Happ \sin \theta -M_\mathrm{S} J_\text{app}T_\text{DL} \theta,
	\end{equation}
where the effect of the SOT is introduced by the last term% and 
%$ g(J_\text{app})=M_\mathrm{S} J_\text{app}T_\text{DL}$
~\cite{spotSuppMat}.
The equilibrium magnetization angle in the center, $\theta_\mathrm{c}$ is found by minimizing 
Eq.~\ref{eq:et}, while edge angle $\theta_\mathrm{e}$ is found by solving
$[E(\theta_\mathrm{e})-E(\theta_\mathrm{c})]/V=0.25D^2A^{-1}$~\cite{spotSuppMat}.
For small SOT and $\vec H_\text{app}$, 
two stable solutions for $\theta_\mathrm{e}$ exist, corresponding to both sample edges.
Above a threshold SOT one solution disappears, 
indicating that the magnetization on one edge is unstable, 
i.e.\ domain nucleation occurs. 
Using numerical methods, the critical current for nucleation $J_\mathrm{c}$ can be calculated easily as a function 
of $D$ (see Fig.~\ref{fig:2} a), black line). 
A good agreement is obtained with micromagnetic simulation for a dot diameter $d=100\;$nm (circles). 
For $D$ tending to zero, the nucleation current tends to the critical current predicted by  
the macrospin model  $J_\mathrm{c}=4.1\times10^{12}\;\text{A/m}^2$~\cite{Lee_APL2013}.
The absence of full quantitative agreement with micromagnetic simulation can be attributed to variations of the
demagnetizing tensor and variations of the magnetization along the $y$-direction due to the curvature of the dot.
Better agreement is obtained when neglecting these effects in a quasi 1D simulation (square dots). 
Note that this nucleation current is actually the threshold current for quasi-DC current pulse. 
	
In the following, we discuss the dynamics of the magnetization switching.  
In Fig.~\ref{fig:2}b) the switching time is shown  as a function of $J_\mathrm{app}>J_\mathrm{c}$. 
With increasing $J_\mathrm{app}$ the switching time decreases rapidly as the DW velocity increases~\cite{Boulle_PRL2013}. 
If $D$ is reduced, the DW propagation is slower, resulting in a larger switching time. 
In the inset we show $J_\text{app}$ versus $1/t_0$ for $D=2\;\text{mJ/m}^2$: 
a linear scaling is observed, 
in qualitative agreement with experiment~\cite{Garello_APL2014}.

Naturally, $t_0$ depends on the dot diameter. This is a key parameter for SOT applications.
The evolution of switching time versus current density for varying dot sizes is shown in 
Fig.~\ref{fig:2} c). When decreasing the diameter from $100\;$nm down to $50\;$nm,
a shift to shorter switching times is observed while a slightly higher onset current is found. 
Similar behavior is 
found when decreasing the size down to $30\;$nm and further down to $25\;$nm. It is, however, important to note 
that the latter two graphs become identical for larger $\Japp$. 
Reducing the size down to $15\;$nm, 
results in a dramatic increase of the threshold current density. 
Moreover, deterministic switching is 
observed in a narrow current density region only. Overall one has indications for three 
different size-dependent
switching regimes. 
In the first regime the switching is covered by nucleation and propagation of a DW 
and the decrease of $t_0$ is mainly caused by a reduced distance for the DW to travel.
In the second regime the switching remains governed by DW propagation. The diameter, however, 
becomes comparable to approximately twice the value of 
$\xi=2A/D\approx 10$\;nm, the characteristic length scale on which canting of the edge \mgn\ is observed. 
In this situation the edge angle due to DMI differs from the ideal infinite case and opposite edges are not completely
independent anymore (see Ref.~\cite{spotSuppMat}). 
While this does not cause coherent rotation yet, it affects the DW motion. 
The coherent regime is
reached at diameters in the range of the DW width $\Delta= \pi\sqrt{A/K_\text{eff}}\approx 14\;$nm. This
explains the significant change in switching behavior for the $15\;$nm dot. 
Note that the switching current at this size is close to the one predicted by macrospin 
simulation ($4.1\times10^{12}\;\text{A/m}^2$). 
It is worth mentioning that while the current density strongly increases with decreasing dot diameter, the current in the
$3\;$nm thick Pt stripe decreases almost linearly, as can be seen from Fig.~\ref{fig:2} d). 
Therefore, the device exhibits favorable scaling behavior and
assuming a $1\;\mathrm{k}\Omega$ resistance for the addressing transistor of 
a $30\;$nm dot, switching in about $300\;$ps, needs only $20\;$fJ for one switching event, 
which is significantly smaller than the energy for perpendicular spin-transfer torque
devices~\cite{Liu_JMMM2014}.

\begin{figure}
		\includegraphics[width=0.85\linewidth]{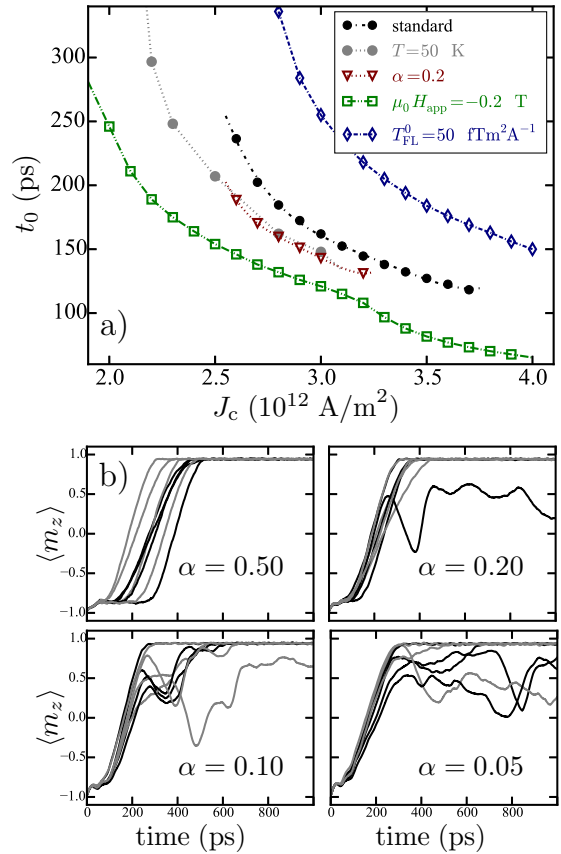}
		\caption{\label{fig:3} (color online)
			a) Switching time as a function of applied current density, varying 
			intrinsic and extrinsic parameters. For the temperature case, 
			the average $t_0$ is plotted. The single event switching time is defined as before, 
			while the average $t_0$ is defined as the time where the probability of stochastical switching reaches $90\%$.
			b) Several switching graphs $\langle m_z\rangle(t)$ for varying damping at $T=50\;$K and
			$\Japp=2.6\;10^{12}\;\mathrm{A/m}^2$. For fixed $\alpha$ variations are only due to temperature fluctuations.
		}
	\end{figure}

Naturally, the threshold current and switching time depend
on several intrinsic as well as extrinsic parameters.
%The influence on the current above which the DW motion
%becomes turbulent will be discussed elsewhere~\cite{spotUP2015}. 
We have studied in detail the influence of the applied field, the damping constant, 
the strength of the field like torque, and temperature. 
The results are shown in Fig.~\ref{fig:3}. Variations in these parameters lead to  
quantitative changes of the nucleation current as well as the switching time. 
In all cases this is mainly attributed to changes in DW velocity;
lower damping increases the wall velocity and so does an in-plane field, as it promotes and 
stabilizes a N{\'e}el type wall. A negative field like torque also stabilizes the DW, 
while a positive 
one destabilizes it, therefore increasing the switching time. 
The edge nucleation/DW propagation mechanism, however, is not affected. 
Most importantly Fig.~\ref{fig:3} a) shows that the mechanism of switching by 
nucleation and propagation is very robust against 
fluctuations due to temperature  (See Ref.~\cite{spotSuppMat} for more details).
The temperature fluctuations strongly decrease the threshold
current (Fig.~\ref{fig:3} b). Temperature effectively lowers the nucleation barrier, 
such that nucleation times get shorter and, 
consequently, the whole switching becomes faster. It has to be pointed out that the 
nucleation still takes place at the same position on the dot edge and the 
overall process remains bipolar with respect to
field and current reversal. This temperature robustness, however, 
strongly relies on the large damping, 
as can be seen from Fig.~\ref{fig:3} b). With decreasing 
$\alpha$ an increasing tendency of oscillations is observed,
such that deterministic switching cannot be guaranteed~\cite{Lee_APL2013}.

To conclude, we have studied the current induced \mgn\ switching of a 
nanomagnet by spin-orbit torques in the presence of Dzyaloshinskii-Moryia 
interaction (DMI).  
%The reversal occurs  by chiral domain nucleation on one edge due to the asymmetric edge magnetization tilting induced by the DMI 
%and the external magnetic field, followed by domain wall propagation. 
The critical switching current strongly decreases with increasing amplitude of DMI  
and we provide a simple analytical model for this dependency.  
This switching mechanism via chiral domain nucleation explains the 
deterministic switching observed experimentally in ultra thin 
Pt/Co/AlO$_x$ even for sub-ns pulses. The switching is mainly introduced by 
the damping like torque, but the field like torque cannot be neglected 
as it strongly influences the switching time.  
Our systematic studies show a change in the reversal mechanism
below diameters of $30\;$nm, while the switching remains deterministic and bipolar. 
However, at $0\;$K the operational window for current densities 
decreases with decreasing dot diameter. 
The influence of temperature on this technological important limit 
will be investigated in the future.
Most importantly, current scalability is maintained.  
Confirming the potential of SOT-MRAM for scalable fast
non-volatile memory application,
our results will help in the design of devices based on this technology. 

This work was funded by the spOt project(318144) of the EC 
under the Seventh Framework Programme.

\appendix
\section{Appendix}

\subsection{General description of sample and technics\label{sec:sample}}
The properties of cobalt films sandwiched between platinum and a insulating
oxide such as AlO are studied. Due to the different substrate and capping materials the inversion symmetry is 
broken along the vertical axis ($\mathcal{O}z$). The magnetization is oriented out-of-plane 
with a strong magnetocrystalline perpendicular anisotropy. In addition to the 
standard micromagnetic energy density, which includes the exchange, the 
magnetocrystalline anisotropy, the Zeeman and the demagnetizing energy, the 
Dzyaloshinskii-Moriya contribution is included according to the following relation:
	\begin{equation}
		E_\text{DM}=D\left[ m_z \frac{\partial m_x}{\partial x}- m_x \frac{\partial m_z}{\partial x} +
			 m_z \frac{\partial m_y}{\partial y}- m_y \frac{\partial m_z}{\partial y}
			%\text{id.}(x \rightarrow y)
			\right]
		\label{eq:1}
	\end{equation}
This expression corresponds to a sample isotropic in the $x$-$y$-plane, where the 
Dzyaloshinskii vector for any in-plane direction $\vec e_u$ is $D(\vec e_z \times \vec e_u)$  with $D$ a uniform 
constant, originating from the symmetry breaking at the $z$-surface~\cite{Thiaville_EPL2012}, 
and $\vec e_u$ being the 
unit vector in direction of an arbitrary $\vec u$. 

Micromagnetic simulations are based on the time integration of Landau-Lifshitz-Gilbert 
equation including the field-like $T_\text{FL}$ 
and damping-like $T_\text{DL}$ spin-orbit torques:
	\begin{eqnarray}
		\frac{\ud \vec m}{\ud t}&=&-\gamma_0 \left[ \vec m \times (\vec H_\text{eff}+\vec H_\text{DM}) \right]+
			\alpha \left( \vec m \times \frac{\ud \vec m}{\ud t} \right)+  \nonumber\\ 
		&&+
		\vec T_\text{FL}+\vec T_\text{DL}
		\label{eq:2}
	\end{eqnarray}
Here $\vec m$ is the unitary vector of the magnetization, $\alpha$ is the Gilbert damping parameter, 
$\gamma_0$ is the product of the vacuum permeability $\mu_0$ and the free electron gyromagnetic 
ratio $\gamma$. For the spin-orbit torques only the first order terms are considered, 
namely $\vec T_\text{FL}=\gamma T^0_\text{FL}J_\text{app}(\vec m\times \vec e_y)$ 
and $\vec T_\text{DL}=\gamma T^0_\text{DL}J_\text{app}\vec m\times(\vec m\times \vec e_y)$, 
where $T^0_\text{FL}=-0.05\;\text{pT}\text{m}^2/\text{A}$ and  
$T^0_\text{DL}=+0.1\;\text{pT}\text{m}^2/\text{A}$ are scalar constants. 
Higher order terms can be found in \textcite{Garello_Nat2013}. 

The appearance of $\vec e_y$ in the definition of the torques is a simplification due to currents in 
$\vec e_x$. In the general case this must be replaced by $\vec e_y \rightarrow \vec e_u=\vec e_z \times \vec e_j$, 
where $\vec e_j$ is the unit vector in the direction of the conventional current, i.e.\ opposite to the electron flow.

\section{\label{sec:lagrange}Lagrange formalism for damping like spin-torque}
The LLG equation in spherical coordinates ($\theta,\phi$) 
can be derived from a Lagrangian by defining a pseudo kinetic energy~\cite{Wegrove_AJP_2012} of the form 
	\begin{equation}
		\Ekin=-\frac{\MS}{\gamma}\dot \phi \cos\theta,
	\end{equation}
where the dot-notation refers to the total time derivative. 
With $U$ as potential energy, the equation of motion is derived from the action
	\begin{equation}
		\mathcal{L}= \Ekin-U,
	\end{equation}
in the typical form of
	\begin{equation}
		\label{eq:llgf}
		\frac{\delta\mathcal{L}}{\delta q}- \udt \frac{\delta\mathcal{L}}{\delta \dot q}=\frac{\delta F}{\delta \dot q},
	\end{equation}
where 
$\delta$ refers to the functional derivative and $q \in \{\theta,\phi\}$. 
Here we add a dissipative term $F$ of the form
	\begin{equation}
		\label{eq:diss}
		F=\frac{\alpha \MS}{2 \gamma}\left( \udt \vec m +\frac{\chi}{\alpha}\vec m\times\vec e_y\right)^2 
	\end{equation}
and $\chi=\gamma T^0_\text{DL} J_\text{app}$.
The square of the first term in parenthesis of Eq.~\ref{eq:diss} results in the standard damping of the LLG equation. 
The square of the second term vanishes when evaluating Eq.~\ref{eq:llgf}, while the mixed term results in the damping-like 
torque.
Using a quasi 1D case with potential energy of the form
	\begin{equation}
		U=A \theta_x^2-D\theta_x-\Keff \cos^2 \theta -\mu_0 \MS \Happ \sin\theta.
	\end{equation}
(here the index $x$ indicates the partial derivative with respect to $x$) and a field 
applied in $x$-direction one derives the equations of motion as.
	\begin{eqnarray}
		\frac{\MS}{\gamma}\dot\phi \sin\theta-\Keff\sin 2 \theta+2 A \theta_{x x}
			+&&\nonumber\\
			+\mu_0 \MS \Happ \cos\theta-\frac{\alpha \MS}{\gamma} \dot\theta +\frac{\MS}{\gamma} \chi \cos\phi&=&0\\
		\frac{\MS}{\gamma}\dot \theta\sin\theta +\frac{\alpha \MS}{\gamma} \dot\phi \sin^2 \theta+&&\nonumber\\
			+\frac{\MS}{\gamma}\chi \sin\phi\sin\theta\cos\theta&=&0
	\end{eqnarray}
 	\begin{figure*}
		\includegraphics[width=0.75\linewidth]{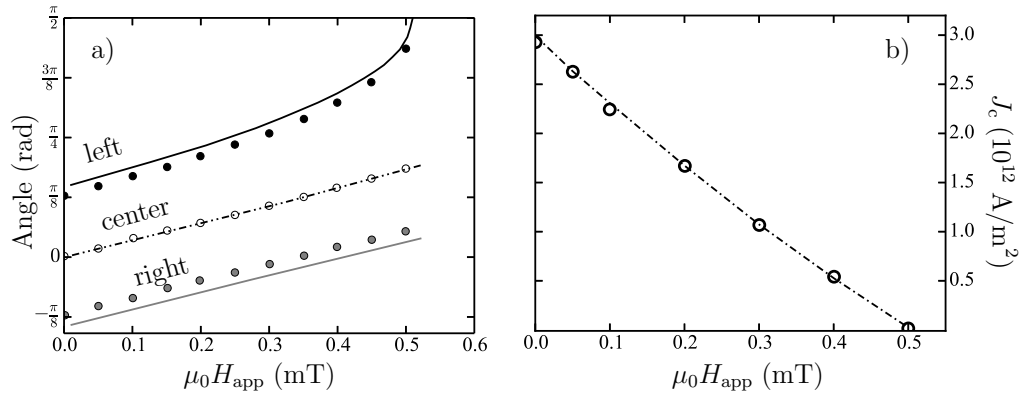}
		\caption{\label{fig:model}
			a) Angle of the magnetization with respect to the $z$-direction as a 
			function of an applied magnetic field in $x$-direction and and zero current. Due to DMI the 
			magnetization at the edges of the infinite long stripe differ from the 
			center. At approximately $0.5\;$T the first edge is fully in-plane. 
			Note, without current and due to symmetry the maximum edge angle is always $\pi/2$.
			b) Analytically calculated critical current $J_\mathrm{c}$, required to 
			destabilize the system, plotted against an in-plane applied magnetic field. 
			In both cases the analytical model (continuous lines) is well reproduced by 
			a micromagnetic simulation (dots) for a stripe of width $200\;$nm ($128$ cells) 
			and a constant effective anisotropy. The other parameters are given in the text.
		}
	\end{figure*}
Note that the DMI drops out when applying the functional derivative. 
In the quasi static case of $\dot\theta=\dot\phi=0$ this simplifies to
	\begin{eqnarray}
		-\Keff\sin 2 \theta+2 A \theta_{x x}
			+\mu_0 \MS \Happ \cos\theta +&&\nonumber\\
			+\frac{\MS}{\gamma} \chi \cos\phi&=&0\\
		\frac{\MS}{\gamma}\chi\sin\phi\sin\theta\cos\theta&=&0
	\end{eqnarray}
The second equations is already consistently fulfilled by setting $\phi=n \pi$, $n\in \mathbb{Z}_0$, 
i.e.\ the quasi 1D case in the $x$-$z$-plane.
It remains to solve
	\begin{equation}
		-\Keff\sin 2 \theta+2 A \theta_{x x}
			+\mu_0 \MS \Happ \cos\theta \pm \frac{\MS}{\gamma}\chi=0
	\end{equation}
Multiplying by $\theta_x$ allows to integrate resulting in
	\begin{equation}
		\label{eq:lgRes}
		A \theta^2_x=-\Keff\cos^2 \theta-\mu_0 \MS \Happ \sin\theta \mp \frac{\MS}{\gamma}\chi \theta+C,
	\end{equation}
where C is the integration constant. This result is identical to Ref.~\cite{Pizzini_PRL2014} except for
the linear term due to the damping like torque, i.e. $g(J_\text{app})\theta$ with 
$g(J_\text{app}) = \MS J_\text{app} T^0_\text{DL}$. 

\section{\label{sec:amodel}Analytical model for the critical current at 0\;K} 
A simplified 
analytical model has been developed, which already exhibits the important 
mechanisms of the behavior observed in experiment. An important ingredient 
for deterministic switching in the analytical model is the presence of DMI. 
The DMI in the presented material system is below the critical value, such 
that no spin spiral is formed. Due to the boundary condition 
\begin{equation}
	\label{eq:bc}
  \frac{D}{2A} \left( \vec e_z \times \vec n \right) \times \vec m = \nabla_{\vec n} \vec m
\end{equation}
and the finite size of the sample, however, 
the \mgn\ is not homogeneous and canting is observed at the edges. Here $\vec n$ is the surface normal and 
$\nabla_{\vec n}$ the normal derivative. 
It has been shown 
recently that the canting angle at the edge as well as in the center of 
the sample can be evaluated using simple energy arguments~\cite{Pizzini_PRL2014}, 
provided that the demagnetizing energy can be approximated by an effective anisotropy. 
This is not always the case~\cite{Meckler_PRB2012}, but for the given parameters it is justified 
as can be seen from \textcite{Rohart_PRB2013}. The derived angles from this simple 
analytical calculation and a micromagnetic simulation that accounts for 
the demagnetization energy by an effective anisotropy, are in very good 
agreement, as can be seen from Fig.~\ref{fig:model} a). As the micromagnetic angle 
is taken from the cell magnetization, a small deviation to the theoretical 
edge value is observed. With decreasing mesh size this deviation decreases. 
The original publication~\cite{Pizzini_PRL2014} calculates a critical magnetic field in 
$z$-direction resulting in the nucleation of a domain. In the present 
case the according term is substituted by a current dependent term (see paragraph~\ref{sec:lagrange}), 
taking account for the dominant damping like torque. At first glance 
this is problematic, as the model considers a static case approaching 
an instability. Moreover, it is 2D in the sense that it assumes $m_y = 0$, 
while a damping like torque initially acts along the $y$-direction. 
For large damping, however,
changes of the magnetization become quasi static and mainly take place in the $x$-$z$-plane.

To provide the full solution for $\theta=\theta(x)$ one needs to integrate Eq.~\ref{eq:lgRes}.
In the center of the sample, i.e.\ far away from the edges, one can assume $\theta_x=0$, such that the DMI
drops out again. The total energy density, hence, has the form
	\begin{equation}
		\label{eq:et}
		\frac{E(\theta)}{V}=-\Keff\cos^2\theta-M_\mathrm{S}\Happ\sin\theta -g(J_\text{app}) \theta
	\end{equation}
and the equilibrium angle is found by minimizing Eq.~\ref{eq:et}.
A second position where one can find a solution is the edge. Here additional information is given 
by the boundary condition Eq.~\ref{eq:bc}. 
%The DMI boundary condition tilts the edge magnetization. The energy contribution
%of DMI can be addressed at a local field, lowering its energy. 
Inserting the boundary conditions into Eq.~\ref{eq:lgRes} results in Eq.~\ref{eq:et} plus an 
additional offset $\Delta E$ of
	\begin{equation}
		\label{eq:de}
		\frac{\Delta E}{V}=\frac{D^2}{4 A}
	\end{equation}
As a result the solution of the edge angle is given by a value $\theta$ that has a $\Delta E$ higher energy than
the minimum energy in the DMI free energy landscape of Eq.~\ref{eq:et}. 
Hence, a stable solution for the edge angle can only be found
if the energy
landscape provides a value of $\Delta E$ above the local minimum, i.e.\ above the solution of the center angle. 
For a small current the edge solution corresponds to a point near the next local maximum. With increasing current
this local maximum decreases and at the critical current $J_\mathrm{c}$, 
it is separated from the minimum 
by exactly $\Delta E$. Eventually, the solution for the magnetization 
angle at the sample edge vanishes for larger currents, 
i.e.\ the current drives the system into instability.
%Note that with zero current and introducing the instability by an in-plane applied field, 
%symmetry provides that the maximum of the energy landscape is fixed at $\pi/2$ and the critical 
%angle for the edge \mgn\ is $\pi/2$ (see Fig.~\ref{fig:model} b)).
%As current introduces a term linear in $\theta$, it breaks symmetry. 
The results of the analytical model
for current driven instabilities are 
plotted as continuous line in Figure~\ref{fig:model} b). The analytical values 
are compared to a simple simulation neglecting variations of the demagnetizing 
field and assuming only an effective anisotropy. The agreement of analytical 
results and simulation are in astonishing agreement. The applicability of 
the model is, however, limited as it neglects temperature. It is shown below 
that temperature has a significant effect on the critical current.

 	\begin{figure}
		\includegraphics[width=0.99\linewidth]{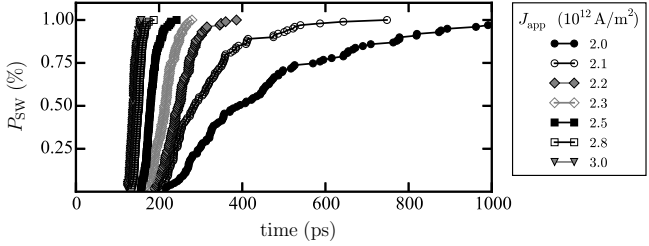}
		\caption{\label{fig:SW}
			Accumulated switching probability as function of time for different current densities at $T=50\;$K.
		}
	\end{figure} 
\section{\label{sec:temp}Thermal fluctuations and switching probabilities}
The temperature was included in the form of a Gaussian distributed 
thermal field $\vec H_\text{th}$, which is added to the effective 
field $\vec H_\text{eff}$. The thermal fluctuations have the following properties~\cite{Brown_PR1963}:
	\begin{equation}
		\langle H_{\text{th},i} (t)\rangle =0
		\label{eq:3}
	\end{equation}
and
	\begin{equation}
		%\langle H_{\text{th},i} (t) H_{\text{th},j} (t')\rangle =\frac{2 \alpha k_\mathrm{B}T}{\gamma_0 \mu_0 \MS V_\text{cell}} \delta_{i j}\delta(t-t')
		\langle H_{\text{th},i} (t) H_{\text{th},j} (t')\rangle =\frac{2 \alpha k_\mathrm{B}T}{ \mu_0  V_\text{cell}} \delta_{i j}\delta\left( \gamma_0 \MS (t-t')\right)
		\label{eq:4}
	\end{equation}
where $k_\mathrm{B}$ is the Boltzmann constant and $V_\text{cell}$ the volume of the discretization cell. 
As temperature results in stochastic \bhvr\ of switching one has to consider switching probabilities. 
This is done by evaluating 100 independent switching events for each parameter set. For each event the 
the switching time $t_0$, i.e.\ the time when $\langle m_z\rangle$ crosses zero is determined. From this the integrated 
probability is calculated. A typical result is shown in Fig.~\ref{fig:SW}. The overall switching time in the presence of 
thermal fluctuations is then defined as the time where the integrated probability reaches 90\%.
Naturally, the switching time decreases with increasing current density. One has to keep in mind, however, that this
data representation does not give information about oscillatory behavior for larger times beyond the (first)
zero crossing of $\langle m_z\rangle$. This graph, hence, can pretend fast switching where actually oscillatory \bhvr\ is
present.

\section{\label{sec:size}Size dependence} 
 	\begin{figure}
		\includegraphics[width=0.99\linewidth]{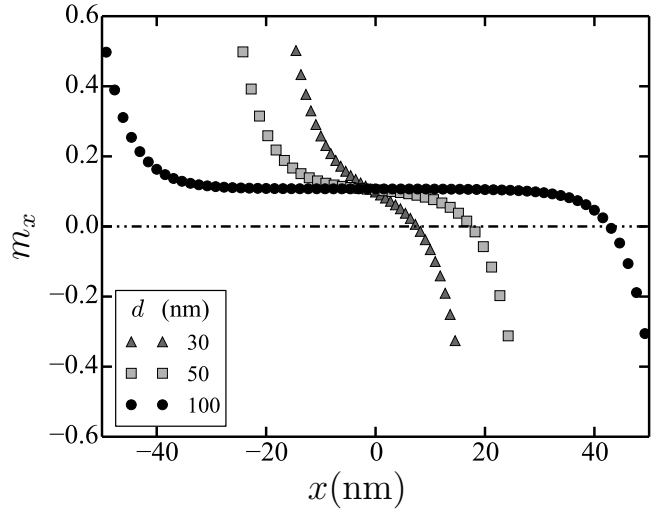}
		\caption{\label{fig:CS}
			Cross section showing $m_x$ for different dot diameters and a $100\;$mT applied field $x$-direction. Within 
			approximately $20\;$nm the canting at the edge
			decays towards a homogeneous magnetization at the dot center.
		}
	\end{figure} 
In addition to the exchange length, the system at hand exhibits  a second important length scale. 
This length is given by the interplay of DMI and exchange interaction, i.e., 
to what extend the non-collinear magnetism 
from the edge penetrates the sample. The according length is given 
by $\xi= 2A/D\approx 10\;$nm. It can be seen from Fig.~\ref{fig:CS} that this length scale
becomes relevant for dot diameters below approximately $40\;$nm. 
For smaller diameters a strictly monotonic change of $m_z$ is observed while larger diameters
present a center plateau. 

%\section{\label{sec:gif}Animations}
%With this supplementary material 2 animations in GIF format are provided. 
%The first animations shows the forward and backward switching of a $100\;$nm dot at $0\;$K.
%After $50\;$ps risetime the current pulse provides a constant $J_\text{app}=2.6 \; 10^{12}\;\mathrm{A}/\mathrm{m}^2$ 
%for $500\;$ps before it falls to zero within $50\;$ps. Without current the magnetization relaxes to equilibrium at
%about $750\;$ps. At this time a current pule in $-x$-direction is injected, such that the system is switched back.
%
%The second animation uses a current pulse with the maximum current applied for $1\;$ns. Most importantly, no external
%field is applied. As a consequence the symmetry is not broken and nucleation continues such that the final 
%state is basically not deterministic. 

\bibliography{../../Mikuszeit_JAB}%
\end{document}